\documentclass{article}
\usepackage[T1]{fontenc} 
\usepackage[utf8]{inputenc} 
\usepackage{ismir,amsmath,cite,url}
\usepackage{graphicx}
\usepackage{color}
\usepackage{booktabs}
\usepackage{amssymb}
\usepackage{xhfill}

\usepackage{lineno}

\title{Transfer Learning and Bias Correction with\\Pre-trained Audio Embeddings}

\threeauthors
  {Changhong Wang\textsuperscript{1}\hspace{-6cm}} {{}\\{}\\{}}
  {Gaël Richard\textsuperscript{1}} {\textsuperscript{1}LTCI, Télécom Paris, Institut Polytechnique de Paris, France\\\textsuperscript{2}Music and Audio Research Laboratory, New York University, USA\\{\tt changhong.wang@telecom-paris.fr}}
  {\hspace{-6cm} Brian McFee\textsuperscript{2}} {{}\\{}\\{}}

\def\authorname{Changhong Wang, Gaël Richard, and Brian McFee}

\begin{document}

\maketitle
\begin{abstract}
Deep neural network models have become the dominant approach to a large variety of tasks within music information retrieval (MIR).
These models generally require large amounts of (annotated) training data to achieve high accuracy.
Because not all applications in MIR have sufficient quantities of training data, it is becoming increasingly common to transfer models across domains.
This approach allows representations derived for one task to be applied to another, and can result in high accuracy with less stringent training data requirements for the downstream task.
However, the properties of pre-trained audio embeddings are not fully understood.
Specifically, and unlike traditionally engineered features, the representations extracted from pre-trained deep networks may embed and propagate biases from the model's training regime.

This work investigates the phenomenon of bias propagation in the context of pre-trained audio representations for the task of instrument recognition.
We first demonstrate that three different pre-trained representations (VGGish, OpenL3, and YAMNet) exhibit comparable performance when constrained to a single dataset, but differ in their ability to generalize across datasets (OpenMIC and IRMAS).
We then investigate dataset identity and genre distribution as potential sources of bias.
Finally, we propose and evaluate post-processing countermeasures to mitigate the effects of bias, and improve generalization across datasets.
\end{abstract}
\section{Introduction}\label{sec:introduction}

\emph{Transfer learning} generally refers to the concept of adapting a model for one task to solve another task.
Often, this is achieved by extracting the internal representation (an \emph{embedding}) of input data from a pre-trained neural network, and providing it as input features to some (often simpler) \emph{downstream} model for the target task.
While this approach is increasingly common and effective, pre-trained embedding models may encode and propagate implicit biases which can have detrimental and disparate population-dependent effects.
Biases have caught wide attention from research fields such as natural language processing (NLP) \cite{garrido2021survey, ntoutsi2020bias, garg2018word}, cognitive science \cite{haselton2015evolution}, and computer vision \cite{tommasi2017deeper}, while in music information retrieval (MIR), bias of pre-trained audio embeddings, is under-explored. 

In this paper, we identify and address the bias of different pre-trained audio embeddings for transfer learning on the task of instrument recognition.
We summarize the contributions as following. (1) We study the within- and cross-domain performance of three pre-trained audio embeddings (VGGish, OpenL3, YAMNet) on two instrument datasets (IRMAS and OpenMIC-2018).
(2) We demonstrate that this approach can propagate bias by producing classifiers which are sensitive to the source domain (dataset).
(3) Based on the performance variation in cross-domain generalization, we investigate dataset identity and genre distribution as potential sources of bias.
(4) We propose a post-processing countermeasure to mitigate unwanted bias in the representation. 
We experiment different bias correction strategies, and analyze the robustness of each pre-trained audio embedding.
The proposed strategies make use of relatively little additional information, and generally produce a modest improvement to cross-domain accuracy for the instrument recognition task.
Our code for all experiments is publicly available\footnote{https://github.com/changhongw/audio-embedding-bias}.

\section{Related work}\label{sec:related}
Pre-trained embeddings are becoming increasingly used in transfer learning for audio-related tasks. 
Choi et~al.~\cite{DBLP:conf/ismir/ChoiFSC17} presented a transfer learning approach for music classification and regression tasks using the internal activations of a pre-trained convolutional network as features. 
The network was trained on the source task of music tagging, and the learned representation was then transferred to five target tasks, including genre classification, vocal/non-vocal classification, emotion prediction, speech/music classification, and acoustic event classification. 
Other well-known audio embedding models include OpenL3~\cite{cramer2019look}, VGGish~\cite{jansen2017large}, and YAMNet\footnote{https://github.com/tensorflow/models/tree/master/research/audioset/\\yamnet}.
The OpenL3 is a 512-dimensional embedding model that results from self-supervised training of the look-listen-learn (L3)-Net for audiovisual correlations.
VGGish (128-dimensional) and YAMNet (1024-dimensional) are both embeddings derived from classification models trained on AudioSet~\cite{jansen2017large}.
Although these embedding models differ in the architecture of the network, source data, and training regime, they have each demonstrated good and comparable generalization performance for a variety of tasks.

Despite that embedding models are normally trained on large amounts of data, it inevitably encodes biases due to the limitation of collected data.
This problem can be especially prevalent in models trained via unsupervised or self-supervised strategies, where there may be no incentive for the model to learn invariances or equivalencies in the data beyond what is required for by the training objective.
As a result, pre-trained embeddings may propagate unwanted biases to downstream tasks.
Different types of biases and bias correction methods are explored in the NLP literature, such as gender~\cite{bolukbasi2016man}, race and religion~\cite{brown2020language}.
A general approach for addressing gender bias in word embedding was proposed by Bolukbasi et al., following three steps: identify bias direction, remove bias by projecting out the bias direction, and equalize pairs~\cite{bolukbasi2016man}.

Besides field-specific biases, dataset bias is a general type of bias that could happen in any application domain.
Tommasi et al.~\cite{tommasi2017deeper} investigated dataset bias in visual recognition with a cross-dataset testbed comprising 12 different datasets.
Ganin et al.~\cite{ganin2016domain} proposed adversarial training for domain adaptation to reduce sensitivity to data drawn from similar but different distributions. 
When detecting depression, a mental health disorder, from speech, Bailey and Plumbley~\cite{bailey2021gender} found that biases in dataset could result in skewed classification performance. 

The approach we take in this paper is most similar to those of Bolukbasi et al.~\cite{bolukbasi2016man} and Ganin et al.~\cite{ganin2016domain}.
While Bolukbasi et al.'s method requires numerous paired examples to identify a subspace which encodes undesirable bias, our proposed method works at the level of collection statistics rather than individual correspondence, and may be easier to apply for audio applications.
Similarly, Ganin et~al.'s method requires adversarial training of the initial model to produce a representation which cannot discriminate well between subsets of data that should be treated equivalently.
Our approach is implemented as a post-processing step, and can be applied to any pre-trained model.
While we focus in this work on dataset identity as a concrete source of bias, we emphasize that the method should be generally applicable to other scenarios in which audio representations exhibit unwanted sensitivity to identifiable attributes.

\section{Methods}\label{sec:methods}
We consider embedding bias from the perspective of \emph{domain adaptation}.
Unlike transfer learning, which relates to the \emph{output} of the model, domain adaptation refers to the behavior of a model (classifier, regressor, etc.) under changes to the distribution of \emph{input data}.
This is closely related to \emph{representation bias}, which is one among many forms of bias known to impact machine learning systems as enumerated by Mehrabi et~al.~\cite{mehrabi2021survey}.
If a classifier is trained on a sample of (labeled) data which is not representative of the target population, then we expect the model to generalize poorly.
The degree to which a pre-trained audio embedding is sensitive to differences between populations of interest---e.g., between a dataset annotated for instrumentation, compared to other collections of music---is therefore of principal interest~\cite{ben2006analysis}.


\subsection{Domain sensitivity}\label{sec:domainadapt}
We investigate the domain sensitivity of three pre-trained embeddings (OpenL3, VGGish, and YAMNet) in transfer learning for the downstream task of instrument recognition.
Each embedding is evaluated in both within-domain and cross-domain setting.
For within-domain evaluation, we train and test the embedding in a single dataset; while for cross-domain case, we investigate the domain adaptation capability of the embedding models across datasets, i.e. training and testing the downstream classifier using data from different datasets.
As a study case, we consider two well-known datasets for instrument recognition, i.e. IRMAS~\cite{bosch2012comparison} and OpenMIC-2018~\cite{humphrey2018openmic} (see Section \ref{subsec:datasets} for dataset details).

\begin{figure}[ht]
    \centering
    \includegraphics[width=\linewidth]{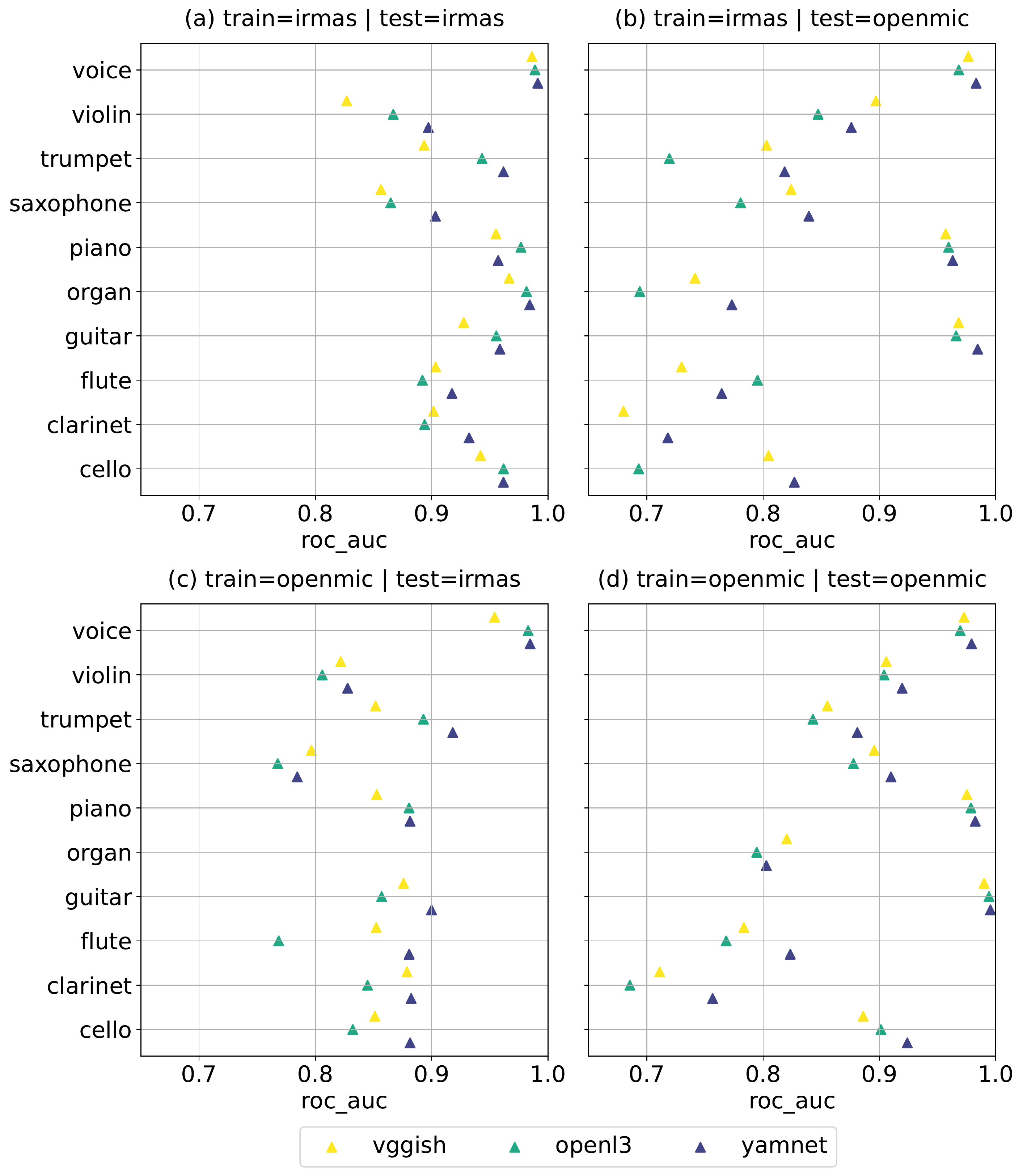}
    \caption{Within-domain (a, d) and cross-domain (b, c) performance of pre-trained audio embeddings (VGGish, OpenL3, and YAMNet) on instrument recognition in the IRMAS and OpenMIC datasets. ROC-AUC refers to area under the receiver operating characteristic curve.}
    \label{fig:within_domain}
\end{figure}

Fig. \ref{fig:within_domain} (a) and (d) visualize the within-domain (IRMAS--IRMAS and OpenMIC--OpenMIC) recognition results in terms of area under the receiver operating characteristic curve (ROC-AUC) using the three embeddings above for each of the ten instrument classes.\footnote{We report AUC because it is invariant to overall class proportions and decision thresholds---which vary between datasets---and thereby allows us to focus on the separating directions identified for each class.}
All three embeddings achieve comparable results, although there is a loose performance ranking of YAMNet $>$ OpenL3 $>$ VGGish for most instrument classes.

\begin{figure*}[ht]
    \centering
    \includegraphics[width=0.9\linewidth]{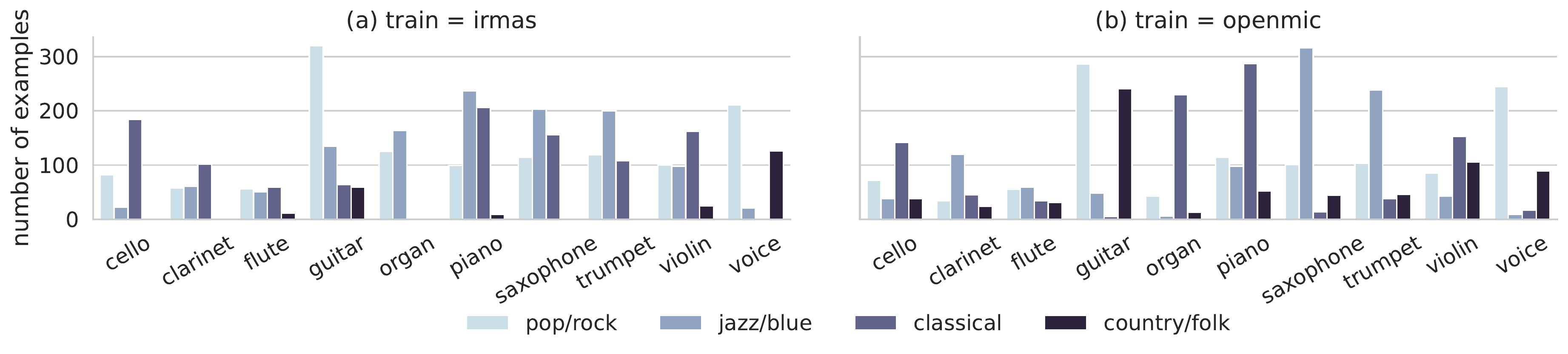}
    \caption{Number of genre examples for each instrument in the training set of IRMAS and OpenMIC datasets. We align the genre labels according to those in the IRMAS dataset: pop/rock, jazz/blue, classical, and country/folk.}
    \label{fig:genre_info}
\end{figure*}


When generalizing across domains, performance degradation happens for both cross-dataset pairs, as shown in Fig. \ref{fig:within_domain} (b) IRMAS--OpenMIC and (c) OpenMIC--IRMAS. 
The performance ranking of the three embedding does not persist either.
Comparing the results when testing on OpenMIC, i.e. (b) and (d), only \emph{voice, piano,} and \emph{guitar} retain close results.
For the remaining instrument classes, all three embeddings exhibit diminished performance.
Similar trends take place in the comparison between (a) and (c) where the test set is IRMAS.

Surprisingly, a dramatic performance drop happens for the \emph{organ} class.
Examining this class in both datasets, we notice a large distribution difference on genre, as shown in Fig.~\ref{fig:genre_info}.
Organ in IRMAS is confined to \emph{pop/rock} and \emph{jazz/blue} genres, suggesting that examples mostly contain electric organ sounds (e.g. Hammond B3).
The distribution of organ in OpenMIC is more balanced, but dominated by classical recordings which are more likely to contain pipe organ than electric.
These differences aside, we generally expect the instrument labels to refer to similar sounds across domains.

\subsection{Quantifying domain bias}\label{sec:quantifybias}

To quantify the effect of domain bias, we first obtain the domain separation direction vector ${\bm w} \in \mathbb{R}^{D}$ by fitting a linear discriminant analysis (LDA) model to discriminate between the OpenMIC and IRMAS datasets in each representation (VGGish, OpenL3, and YAMNet).
${\bm v}_{k} \in \mathbb{R}^{D}$ is the instrument separation direction vector, i.e. the coefficient vector of the trained downstream classifier, for the $k$-th instrument. 
$k=1,2,...,K$ is the instrument class index and $D$ is the dimension of pre-trained embedding.
We measure the correlation between the domain separation and downstream classification using the cosine similarity between ${\bm w}$ and ${\bm v}_{k}$:
\begin{equation}
    {\bm c}_{k}({\bm w}, {\bm v}) = \frac{\langle{\bm w}, {\bm v}_{k} \rangle}{\|{\bm w}\|\times\|{\bm v}_{k}\|}
\end{equation}
Large (in magnitude) ${\bm c}_k$ indicates that the instrument classifier is sensitive to dataset identity.

Fig. \ref{fig:cosine_similarity_perclass} top shows the absolute correlation value for each instrument class, when the classifier is trained on the training set of (a) IRMAS and (b) OpenMIC dataset, respectively.
The mean correlation value over all instruments for each embedding is displayed in the legend.
It clearly shows that YAMNet is the least sensitive to dataset bias; OpenL3 is also relatively stable while VGGish is the most sensitive to dataset bias.
The relatively large correlation value for the organ class matches our analysis in Section~\ref{sec:domainadapt} that genre distribution might be also a potential source of bias (see Fig.~\ref{fig:genre_info}).
Although the sensitivity of different embeddings to dataset bias are different, bias cannot generally be removed by simply using different pre-trained embeddings.
As we will demonstrate, explicitly correcting for dataset bias can potentially improve domain adaptation performance for each choice of embedding.


\subsection{Bias correction}\label{sec:debias}
To mitigate domain bias, we propose a post-processing countermeasure on the pre-trained embeddings which does not interact with the training process of embeddings.
Importantly, the proposed method requires only samples of data which should behave similarly for the downstream task, but it does not require these samples to be \emph{labeled} for the downstream task.

Continuing our instrument classification example, given that both datasets contain examples from each of the instrument categories of interest, we should expect that a well-formed linear classifier should behave independently of the domain from which data is drawn.
Concretely, this means that the separating direction ${\bm v}_k$ should be orthogonal to any direction ${\bm w}$ which separates the two datasets in the embedding space, resulting in ${\bm c}_k({\bm w}, {\bm v}_k) = 0$.
While Ganin et al.~\cite{ganin2016domain} use this intuition to adversarially train the representation, this approach is impractical when using pre-trained embeddings which are presumed to be fixed in advance.
Instead, we approach this problem by post-processing the embedding to project out the direction ${\bm w}$ which separates domains that should be indistinguishable for the downstream task.

Concretely, if ${\bm w} \in \mathbb{R}^D$ is the domain-separating direction (normalized to unit length, $\|{\bm w}\|=1$), we project this dimension out of the space by applying the following transformation to input data ${\bm x} \in \mathbb{R}^D$:
\begin{equation}\label{eq:globaldebias}
    {\bm x}_{\rm P} :=  \left(\mathbf{I} - {\bm w}{\bm w}^\mathsf{T}\right){\bm x}
\end{equation}
where $\mathbf{I}$ is the $D\times D$ identity matrix.
The new embedding ${\bm x}_{\rm P}$ is the input to the classifier.

\subsection{Multiple bias correction}
While the above strategy is defined for binary bias correction, e.g. where there are two domains to be reconciled, it does generalize to more complex settings.
In the instrument recognition example, we may also consider differences between genres across datasets as a source of bias.
Even if two datasets both consist of examples in the same genre categories, this does not necessarily mean that the genre terms are used consistently between datasets.

To consider the influence of genre distribution, we propose also \emph{multiple bias correction}, where we extract the dataset separation direction in the genre subspace.
That is, for each pair of matched genre labels, e.g. pop/rock in IRMAS and pop/rock in OpenMIC, we fit a binary LDA to separate them.
Then for each genre category $g = 1, 2, \dots G$ (for $G \geq 1$ genres), we obtain a dataset separation direction vector ${\bm w}_g$ which depends only on examples from genre $g$.
Collecting all ${\bm w}_g$ into a matrix $W \in \mathbb{R}^{D\times G}$ defines a basis for a subspace of the embedding of dimension at most $G$.
Note that $W$ may not be an orthogonal basis, as different ${\bm w}_g$ may correlate with each other.

We therefore derive an orthogonal basis by factorizing $W$ via the reduced singular vector decomposition (SVD):
\begin{equation}
    W = U\Sigma V^\mathsf{T}
\end{equation}
where $\Sigma$ is a $G\times G$ diagonal matrix of singular values, and $U\in\mathbb{R}^{D\times G}$ and $V\in \mathbb{R}^{D\times G}$ are the left- and right-singular vectors.
We use the right singular vectors as an orthogonal basis for the domain-separating subspace, resulting in the following generalization of Eq.~\eqref{eq:globaldebias}:
\begin{equation}
    {\bm x}_{\rm P} := \left(\mathbf{I} - VV^\mathsf{T}\right){\bm x} \label{eq:genredebias}
\end{equation}

In applying Eq.~\eqref{eq:genredebias}, it is important to verify that $W$ is full rank ($G$), e.g. by verifying that all singular values $\Sigma$ are sufficiently large, as Eq.~\eqref{eq:genredebias} would otherwise remove a larger than necessary subspace from the representation.
In all cases studied in this work, $W$ was full rank.

\subsection{Nonlinear bias correction}
The above methods are based on two assumptions: 1) that the downstream model will be linear, and 2) that the domains are linearly separable.
These assumptions may be restrictive in practice, so we generalize the method above by transforming the embeddings to a higher dimensional space using kernel methods.
While both logistic regression and linear discriminant analysis support kernel generalizations~\cite{hastie2009elements}, the subspace projection method described above is less directly adaptable.\footnote{One could achieve a similar effect by adding a linear constraint $\langle {\bm w}, {\bm v}\rangle = 0$ to the logistic regression problem, but this would require a custom solver and limit the general utility of the approach.}

Instead of using implicit kernel representations, we will use approximate, i.e. explicit kernel approximation.
That is, instead of replacing inner products $\langle {\bm w}, {\bm v} \rangle$ by nonlinear kernel function calculations $k({\bm w}, {\bm v})$, we apply an explicit nonlinear transformation $f : \mathbb{R}^D \rightarrow \mathbb{R}^{D'}$ such that
\begin{equation}
    \langle f({\bm w}), f({\bm v}) \rangle \approx k({\bm w}, {\bm v})
\end{equation}
We then apply the previously defined bias correction methods on the transformed data $f({\bm w})$, which results in projecting out the dataset-separating direction(s) after applying $f$ but prior to fitting the downstream (instrument) classifiers.

There are several choices to be made here when selecting the kernel $k$ and the approximating map $f$.
In this work, we use a standard radial basis function (Gaussian) kernel and the ``random Fourier features'' approximation method~\cite{rahimi2007random}.
However, we note that other choices (e.g., the Nystr\"om method) are readily available in scikit-learn~\cite{scikit-learn}, and may work just as well.

In total, we have four bias-correction strategies: linear bias correction (\emph{LDA}), linear multiple bias correction (\emph{mLDA}), nonlinear bias correction in the kernelized embedding space (\emph{KLDA}), nonlinear multiple bias correction in the kernelized embedding space (\emph{mKLDA}).

\section{Experiments}\label{sec:experiments}

\begin{figure*}[t]
    \centering
    \includegraphics[width=0.85\linewidth]{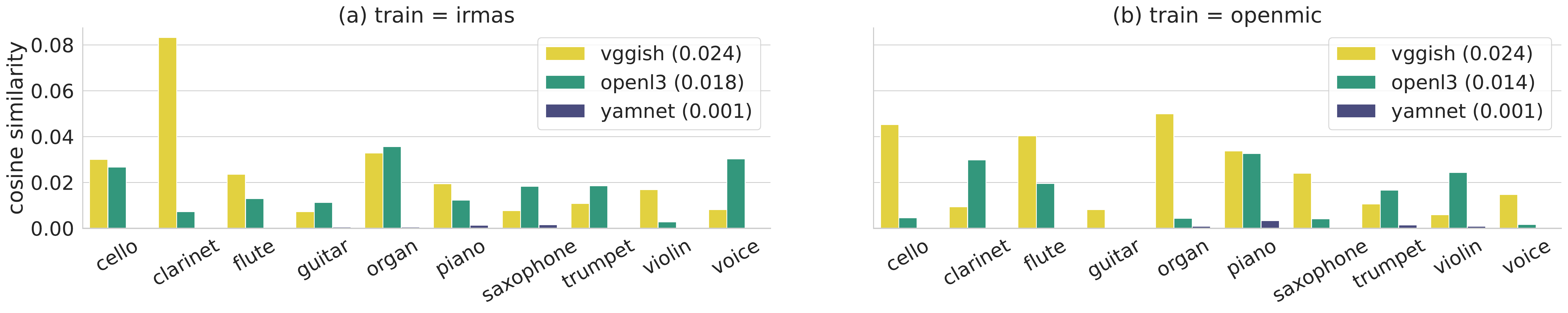}\\
    \includegraphics[width=0.85\linewidth]{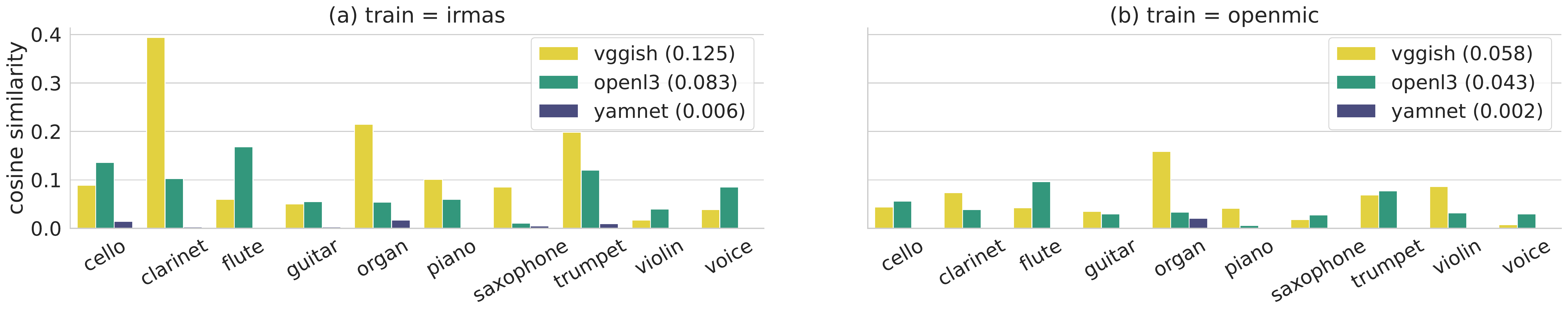}\\
    \includegraphics[width=0.85\linewidth]{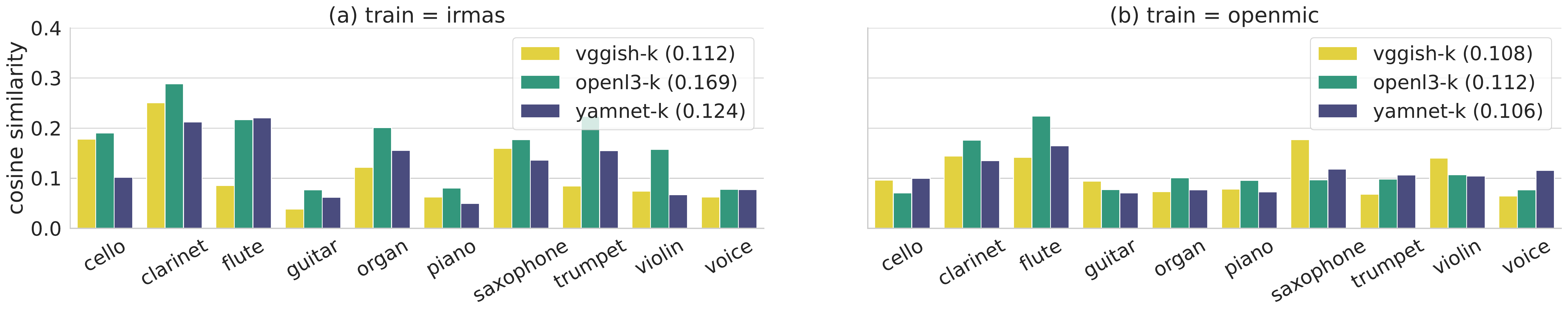}\\
    \caption{Correlation between domain separation and instrument classification for each instrument in the IRMAS and OpenMIC training set. 
    \textit{(Top)}: correlation in the original embedding space with domain separation direction extracted using only dataset identity; \textit{(middle)}: correlation in the original embedding space with domain separation direction extracted class-wise; \textit{(bottom)}: same as middle but in the kernelized embedding space. Mean value is given in parentheses. }
    \label{fig:cosine_similarity_perclass}
\end{figure*}

\begin{table*}[t]\small{
    \centering
    \begin{tabular}{p{2.5cm}p{.1cm}p{.1cm}p{1.8cm}p{1.8cm}p{.1cm}p{.1cm}p{1.8cm}p{1.8cm}}
    \toprule              
     & \multicolumn{4}{c}{Global bias correction}  & \multicolumn{4}{c}{Class-wise bias correction}  \\  
    Debiasing method & \multicolumn{2}{c}{Within-domain} & \multicolumn{2}{c}{Cross-domain} & \multicolumn{2}{c}{Within-domain}  & \multicolumn{2}{c}{Cross-domain}\\
     & \multicolumn{1}{l}{IR-IR} & \multicolumn{1}{l}{OP-OP} & \multicolumn{1}{c}{OP-IR} & \multicolumn{1}{c}{IR-OP} & \multicolumn{1}{l}{IR-IR} & \multicolumn{1}{l}{OP-OP} & \multicolumn{1}{c}{OP-IR} & \multicolumn{1}{c}{IR-OP} \\\midrule
    \textit{VGGish} & \textit{91.6} & \textit{87.95} & \textit{\underline{82.82}} & \textit{\underline{83.81}} & \textit{91.60} & \textit{87.95} & \textit{\underline{82.82}} & \textit{\underline{83.81}} \\
    VGGish-LDA & 91.60 & 87.99 & 82.99 (\textbf{+0.18}) & 83.82 (0.0) & 91.60 & 87.94 & 82.93 (\textbf{+0.12}) & 83.85 (+0.03) \\
    VGGish-mLDA & 91.45 & 87.98 & 82.70 (-0.11) & 83.30 (-0.51) & 91.56 & 87.87 & 83.13 (\textbf{+0.31}) & 83.66 (-0.16) \\
    VGGish-K & 92.24 & 88.08 & 82.57 (-0.25) & 83.67 (-0.14) & 92.24 & 88.08 & 82.57 (-0.25) & 83.67 (-0.14) \\
    VGGish-KLDA & 92.24 & 88.08 & 82.58 (-0.24) & 83.67 (-0.14) & 92.22 & 88.07 & 82.70 (-0.12) & 83.78 (-0.04) \\
    VGGish-mKLDA & 92.22 & 88.15 & 82.42 (-0.39) & 83.70 (-0.11) & 92.26 & 88.08 & 82.70 (-0.11) & 83.76 (-0.05) \\
    \midrule
    \textit{OpenL3} & \textit{93.26} & \textit{87.16} & \textit{\underline{80.56}} & \textit{\underline{80.13}} & \textit{93.26} & \textit{87.16} & \textit{\underline{80.56}} & \textit{\underline{80.13}} \\
    OpenL3-LDA   & 93.26 & 87.16 & 80.56 (+0.01) & 80.15 (+0.02) & 93.24 & 87.18 & 80.59 (+0.04) & 80.38 (\textbf{+0.26}) \\
    OpenL3-mLDA  & 93.11 & 87.16 & 80.67 (\textbf{+0.12}) & 79.93 (-0.20) & 93.09 & 87.23 & 80.57 (+0.02) & 80.62 (\textbf{+0.50}) \\
    OpenL3-K     & 93.89 & 87.91 & 79.46 (-1.09) & 81.23 (\textbf{+1.11}) & 93.89 & 87.91 & 79.46 (-1.09) & 81.23 (\textbf{+1.11}) \\
    OpenL3-KLDA  & 93.89 & 87.84 & 79.03 (-1.53) & 81.23 (\textbf{+1.11}) & 93.96 & 87.91 & 79.99 (-0.57) & 81.79 (\textbf{+1.66}) \\
    OpenL3-mKLDA & 93.88 & 87.88 & 79.56 (-1.00) & 81.20 (\textbf{+1.07}) & 94.04 & 87.83 & 79.97 (-0.59) & 81.32 (\textbf{+1.19}) \\
    \midrule
    \textit{YAMNet} & \textit{94.65} & \textit{89.74} & \textit{\underline{85.01}} & \textit{\underline{85.47}} & \textit{94.65} & \textit{89.74} & \textit{\underline{85.01}} & \textit{\underline{85.47}} \\
    YAMNet-LDA  & 94.65 & 89.74 & 85.01 (0.0)   & 85.47 (0.0)   & 94.65 & 89.74 & 85.02 (0.0)   & 85.47 (0.0) \\
    YAMNet-mLDA & 94.65 & 89.74 & 85.01 (0.0)   & 85.47 (0.0)   & 94.65 & 89.74 & 85.02 (0.0)   & 85.46 (0.0) \\
    YAMNet-K    & 93.83 & 89.24 & 85.87 (\textbf{+0.86}) & 84.56 (-0.91) & 93.83 & 89.24 & 85.87 (\textbf{+0.86}) & 84.56 (-0.91) \\
    YAMNet-KLDA & 93.83 & 89.23 & 85.87 (\textbf{+0.86}) & 84.56 (-0.91) & 93.63 & 89.24 & 86.00 (\textbf{+0.99}) & 84.76 (-0.70) \\
    YAMNet-mKLDA& 93.79 & 89.19 & 85.72 (\textbf{+0.71}) & 84.43 (-1.04) & 93.79 & 89.34 & 85.53 (\textbf{+0.51}) & 84.60 (-0.87) \\
    \bottomrule
  \end{tabular}}
\caption{Mean ROC-AUC (\%) of global bias correction and class-wise bias correction on instrument classification in IRMAS (IR) and OpenMIC (OP) datasets. VGGish, OpenL3, and YAMNet (in \textit{italic}) refers to the original embedding; the other cases, i.e. with -LDA, -mLDA, -LDA, and -mKLDA, correspond to linear, linear-multiple, nonlinear, and nonlinear-multiple debiasing strategies; cases with -K are the kernelized embeddings. Values in parenthesis are the performance boost (>0.1 are \textbf{bolded}) or degradation as compared to the original embedding (the closest \textit{\underline{underlined}} above).}
\label{tab:irmas-openmic}
\end{table*}

\subsection{Datasets and experimental details}\label{subsec:datasets}
The datasets we use are two well-known datasets with instrument annotations, IRMAS~\cite{bosch2012comparison} and  OpenMIC-2018~\cite{humphrey2018openmic}.
The former comprises 20,000 examples of 10-second excerpts, partially labeled for the presence or absence of 20 instrument classes; and the latter contains 6705 audio files of 3-second clips, of which only the predominant instrument were annotated.
Since there are 20 instrument classes in the OpenMIC dataset and 11 in the IRMAS, we focus only on the 10 mutual classes: \emph{cello, clarinet, flute, guitar, organ, piano, saxophone, trumpet, violin,} and \emph{voice}.
For the sake of consistency, \emph{electric guitar} and \emph{acoustic guitar} in the IRMAS dataset have been merged into a single class: guitar.

To investigate the impact of genre, we also align the genres in the two datasets. 
Each audio sample in the IRMAS dataset is labeled with one of the five genres: \emph{pop/rock, jazz/blue, classical, country/folk,} and \emph{latin/soul}; while samples in the OpenMIC datase has multiple labels from around 130 genres.
We consider four genres (pop/rock, jazz/blue, classical, country/folk) as the latin/soul genre has few examples in both datasets. 
The genre labels of the OpenMIC dataset are merged into those of IRMAS with name intersections. 
For example, we merge the genres---\emph{Rock, Loud-Rock,  Noise-Rock, Psych-Rock}, et. al.---in OpenMIC into one genre label: pop/rock.
Multiple genre labels in the OpenMIC dataset are reduced to a single label by the first activation from the four considered genres or the first of the original labels otherwise.

With the embedding features extracted using pre-trained VGGish, OpenL3, and YAMNet models, we train a logistic regression classifier for each instrument class using IRMAS and OpenMIC training data.
The input to the classifier is the mean frame embedding of each audio example.
For OpenMIC dataset, we follow the train-test split in~\cite{humphrey2018openmic}, with a ratio of 3:1.
For fair comparisons, we create a new partition with the same train-test ratio on IRMAS dataset which takes into account of the class-balance and non-track overlap between training and test sets.
To focus only on distribution shift, we use the same number of samples per class for both datasets during training, following the lower one.

For the nonlinear method, we first standardize the embedding features using z-score normalization with the training-set statistics.
Then we approximate the kernels for the embeddings with a fixed dimension $D'$ of four times the dimension ($D$) of the original embeddings.
Finally, we tune the hyper-parameter for the logistic regression classifier, i.e. the inverse of regularization strength $C$, by cross-validation with a grid of $10^{-8:1:4}$.

\subsection{Results}\label{subsec:results}

Table.~\ref{tab:irmas-openmic} lists the instrument classification performance of the debiasing methods discussed in Section \ref{sec:methods} in terms of mean ROC-AUC over the 10 instrument classes. 
To compare the performance of using only dataset identity as additional information and that uses also class-labels, we present two sets of results: \emph{global bias correction} and \emph{class-wise bias correction}.
We first present some observations that are common to both cases and then discuss the comparison.
For the original embeddings (in \emph{italic}), large performance drop shows for all cross-domain cases.
OpenL3 is most sensitive to distribution shift, with a drop of 12.7\% and 7\% when testing on IRMAS and OpenMIC dataset, respectively.
Yet, from the cosine similarity values in Fig. \ref{fig:cosine_similarity_perclass} top and middle, OpenL3 does not embed the most domain bias.
This may indicate that for the task at hand, other more significant distribution shifts that OpenL3 is sensitive to may exist.
For all embeddings, projecting to the higher dimensional space (debiasing methods with ``K'') almost never substantially hurts the within-domain performance and sometimes improves the performance.

Interestingly, when comparing linear debiasing (``-LDA'' and ``-mLDA'') with nonlinear debiasing (``-KLDA'' and ``-mKLDA'') for all embeddings, we find that kernelization does not help for VGGish while YAMNet only works in the kernelized embedding space.
This explains the relative increase of cosine similarity values for YAMNet after kernelization as compared to the other two embeddings (see Fig. \ref{fig:cosine_similarity_perclass} bottom).
Both linear and nonlinear debiasing exhibit performance improvement for OpenL3. 
In terms of global bias correction, almost no improvement for VGGish except LDA for OpenMIC->IRMAS; OpenL3 yields some boost for both cross-domain cases. 
YAMNet improves the results only for OpenMIC->IRMAS. 
It is expected that the class-wise bias correction achieves better performance as we extract the domain bias for the target instrument exactly.
This is also verified by the more noticeable cosine similarity values in the middle subfigure as compared to the top of Fig. \ref{fig:cosine_similarity_perclass}.
VGGish and OpenL3 yields slight improvement for most linear debiasing.
All nonlinear debiasing improves the results of OpenL3 for IRMAS->OpenMIC and YAMNet for OpenMIC->IRMAS.
Although the overall improvement is not significant, we observe large improvements for some instrument classes.

\section{Discussion}
We notice two important factors for transfer learning with pre-trained audio embeddings: the training regime of the embeddings, and the class vocabulary alignment between the source task and downstream task.

The better generalization performance of YAMNet and VGGish in a transfer setting may be attributed to their training regime. 
YAMNet and VGGish are derived from supervised training while OpenL3 is from self-supervised training and more prone to overfitting a domain.
As a result, YAMNet and VGGish have both been incentivized to learn invariances within specific categories (including musical instrumentation), while OpenL3 has no such incentive as it is only designed to predict audio-visual correspondence.
Moreover, YAMNet was specifically trained for sound classification using a vocabluary that broadly subsumes that of our downstream task (instrumentation).
This likely contributes to its high performance and cross-domain stability overall.

The class vocabulary alignment is related to \emph{label shift}, an under-explored type of distribution shift in the domain-adaptation field \cite{azizzadenesheli2019regularized}.
The labelling scheme difference between the two datasets complicates the debiasing as the IRMAS dataset only contains labels for the predominant instrument while all active instruments are annotated in the OpenMIC dataset.
Aligning these two datasets is nontrivial as it involves label shift besides \emph{covariate shift}.
We propose multiple-bias correction, i.e. debiasing in the genre subpsace, to deal with this problem.
Yet, it does not resolve the conditional probability shift that happens due to unbalanced relationships between instrumentation and genre, e.g. the strong dependence between organ and pop/jazz in IRMAS, and in this specific case an argument could be made that the classification task is closer to transfer learning than domain adaptation.

A notable limitation of the presented experiments is the small amount of functional data.
Although OpenMIC dataset is relatively large with 14915 samples for training, only a small portion is actually used in the binary classification of each instrument.
After equalizing the number of samples per class in both datasets, there are only 288, 221, 177, 578, 290, 551, 476, 427, 385, and 358 samples for the 10 instrument classes in the binary classification.
Most classes have number of samples less than the dimension of OpenL3 (512) and all of them are below that of YAMNet (1024). 

\section{Conclusion}
The method proposed in this work addresses one specific form of bias that can arise in transfer learning scenarios.
Correctly applying this method requires identifying subsets of data that should be treated equivalently, i.e., be indistinguishable under the chosen representation.
We stress that this notion of equivalence ultimately depends on the choice of the downstream task, and caution should be exercised when identifying populations to treat as equivalent.
For the case study presented here---domain adaptation and instrument recognition---we argue that the downstream task ought to be generally independent of the source domain, though we recognize that this will not always be true in practice.
We therefore urge practitioners to critically investigate all assumptions of equivalence and independence when applying bias correction methods.


\section{Acknowledgments}
This work was partly funded by the European Union (ERC, HI-Audio, 101052978). Views and opinions expressed are however those of the author(s) only and do not necessarily reflect those of the European Union or the European Research Council. Neither the European Union nor the granting authority can be held responsible for them.
We thank Michel Olvera for the discussions on domain adaptation.

\bibliography{ISMIRtemplate}

\begin{thebibliography}{10}
\providecommand{\url}[1]{#1}
\csname url@samestyle\endcsname
\providecommand{\newblock}{\relax}
\providecommand{\bibinfo}[2]{#2}
\providecommand{\BIBentrySTDinterwordspacing}{\spaceskip=0pt\relax}
\providecommand{\BIBentryALTinterwordstretchfactor}{4}
\providecommand{\BIBentryALTinterwordspacing}{\spaceskip=\fontdimen2\font plus
\BIBentryALTinterwordstretchfactor\fontdimen3\font minus
  \fontdimen4\font\relax}
\providecommand{\BIBforeignlanguage}[2]{{%
\expandafter\ifx\csname l@#1\endcsname\relax
\typeout{** WARNING: IEEEtran.bst: No hyphenation pattern has been}%
\typeout{** loaded for the language `#1'. Using the pattern for}%
\typeout{** the default language instead.}%
\else
\language=\csname l@#1\endcsname
\fi
#2}}
\providecommand{\BIBdecl}{\relax}
\BIBdecl

\bibitem{garrido2021survey}
I.~Garrido-Mu{\~n}oz, A.~Montejo-R{\'a}ez, F.~Mart{\'\i}nez-Santiago, and L.~A.
  Ure{\~n}a-L{\'o}pez, ``A survey on bias in deep {NLP},'' \emph{Applied
  Sciences}, vol.~11, no.~7, p. 3184, 2021.

\bibitem{ntoutsi2020bias}
E.~Ntoutsi, P.~Fafalios, U.~Gadiraju, V.~Iosifidis, W.~Nejdl, M.-E. Vidal,
  S.~Ruggieri, F.~Turini, S.~Papadopoulos, E.~Krasanakis \emph{et~al.}, ``Bias
  in data-driven artificial intelligence systems—an introductory survey,''
  \emph{Wiley Interdisciplinary Reviews: Data Mining and Knowledge Discovery},
  vol.~10, no.~3, p. e1356, 2020.

\bibitem{garg2018word}
N.~Garg, L.~Schiebinger, D.~Jurafsky, and J.~Zou, ``Word embeddings quantify
  100 years of gender and ethnic stereotypes,'' \emph{Proceedings of the
  National Academy of Sciences}, vol. 115, no.~16, pp. E3635--E3644, 2018.

\bibitem{haselton2015evolution}
M.~G. Haselton, D.~Nettle, and D.~R. Murray, ``The evolution of cognitive
  bias,'' \emph{The handbook of evolutionary psychology}, pp. 1--20, 2015.

\bibitem{tommasi2017deeper}
T.~Tommasi, N.~Patricia, B.~Caputo, and T.~Tuytelaars, ``A deeper look at
  dataset bias,'' in \emph{Domain adaptation in computer vision
  applications}.\hskip 1em plus 0.5em minus 0.4em\relax Springer, 2017, pp.
  37--55.

\bibitem{DBLP:conf/ismir/ChoiFSC17}
K.~Choi, G.~Fazekas, M.~B. Sandler, and K.~Cho, ``Transfer learning for music
  classification and regression tasks,'' in \emph{Proceedings of the
  International Society for Music Information Retrieval Conference ({ISMIR})},
  2017, pp. 141--149.

\bibitem{cramer2019look}
A.~Cramer, H.~Wu, J.~Salamon, and J.~Bello, ``Look, listen, and learn more:
  Design choices for deep audio embeddings,'' in \emph{Proceedings of the IEEE
  International Conference on Acoustics, Speech and Signal Processing
  (ICASSP)}, 2019, pp. 3852--3856.

\bibitem{jansen2017large}
A.~Jansen, J.~F. Gemmeke, D.~P. Ellis, X.~Liu, W.~Lawrence, and D.~Freedman,
  ``Large-scale audio event discovery in one million {Y}ou{T}ube videos,'' in
  \emph{Proceedings of the IEEE International Conference on Acoustics, Speech
  and Signal Processing (ICASSP)}, 2017, pp. 786--790.

\bibitem{bolukbasi2016man}
T.~Bolukbasi, K.-W. Chang, J.~Y. Zou, V.~Saligrama, and A.~T. Kalai, ``Man is
  to computer programmer as woman is to homemaker? debiasing word embeddings,''
  \emph{Advances in neural information processing systems}, vol.~29, 2016.

\bibitem{brown2020language}
T.~Brown, B.~Mann, N.~Ryder, M.~Subbiah, J.~D. Kaplan, P.~Dhariwal,
  A.~Neelakantan, P.~Shyam, G.~Sastry, A.~Askell \emph{et~al.}, ``Language
  models are few-shot learners,'' \emph{Advances in neural information
  processing systems}, vol.~33, pp. 1877--1901, 2020.

\bibitem{ganin2016domain}
Y.~Ganin, E.~Ustinova, H.~Ajakan, P.~Germain, H.~Larochelle, F.~Laviolette,
  M.~Marchand, and V.~Lempitsky, ``Domain-adversarial training of neural
  networks,'' \emph{The journal of machine learning research}, vol.~17, no.~1,
  pp. 2096--2030, 2016.

\bibitem{bailey2021gender}
A.~Bailey and M.~D. Plumbley, ``Gender bias in depression detection using audio
  features,'' in \emph{Proceedings of the IEEE European Signal Processing
  Conference (EUSIPCO)}, 2021, pp. 596--600.

\bibitem{mehrabi2021survey}
N.~Mehrabi, F.~Morstatter, N.~Saxena, K.~Lerman, and A.~Galstyan, ``A survey on
  bias and fairness in machine learning,'' \emph{ACM Computing Surveys (CSUR)},
  vol.~54, no.~6, pp. 1--35, 2021.

\bibitem{ben2006analysis}
S.~Ben-David, J.~Blitzer, K.~Crammer, and F.~Pereira, ``Analysis of
  representations for domain adaptation,'' \emph{Advances in neural information
  processing systems}, vol.~19, 2006.

\bibitem{bosch2012comparison}
J.~J. Bosch, J.~Janer, F.~Fuhrmann, and P.~Herrera, ``A comparison of sound
  segregation techniques for predominant instrument recognition in musical
  audio signals.'' in \emph{Proceedings of the International Society for Music
  Information Retrieval Conference ({ISMIR})}, 2012, pp. 559--564.

\bibitem{humphrey2018openmic}
E.~Humphrey, S.~Durand, and B.~McFee, ``Openmic-2018: An open data-set for
  multiple instrument recognition.'' in \emph{Proceedings of the International
  Society for Music Information Retrieval Conference ({ISMIR})}, 2018, pp.
  438--444.

\bibitem{hastie2009elements}
T.~Hastie, R.~Tibshirani, J.~H. Friedman, and J.~H. Friedman, \emph{The
  elements of statistical learning: data mining, inference, and
  prediction}.\hskip 1em plus 0.5em minus 0.4em\relax Springer, 2009.

\bibitem{rahimi2007random}
A.~Rahimi and B.~Recht, ``Random features for large-scale kernel machines,''
  \emph{Advances in neural information processing systems}, vol.~20, 2007.

\bibitem{scikit-learn}
F.~Pedregosa, G.~Varoquaux, A.~Gramfort, V.~Michel, B.~Thirion, O.~Grisel,
  M.~Blondel, P.~Prettenhofer, R.~Weiss, V.~Dubourg, J.~Vanderplas, A.~Passos,
  D.~Cournapeau, M.~Brucher, M.~Perrot, and E.~Duchesnay, ``Scikit-learn:
  Machine learning in {P}ython,'' \emph{Journal of Machine Learning Research},
  vol.~12, pp. 2825--2830, 2011.

\bibitem{azizzadenesheli2019regularized}
K.~Azizzadenesheli, A.~Liu, F.~Yang, and A.~Anandkumar, ``Regularized learning
  for domain adaptation under label shifts,'' in \emph{Proceedings of the
  International Conference on Learning Representations (ICLR)}, 2019.

\end{thebibliography}


%
%
%
%
%

\end{document}